\documentclass[cameraready]{Interspeech}

\title{SpAArSIST: Sparsified AASIST for Efficient and Reliable Anti-Spoofing}

\author[orcid=0000-0002-4717-1910,correspondingauthor]{Anton}{Firc}
\author[orcid=0009-0000-5722-0571]{Vojtěch}{Staněk}
\author[orcid=0009-0001-1870-7704]{Zbyněk}{Lička}
\author[orcid=0000-0002-9009-2193]{Kamil}{Malinka}
\author[orcid=0000-0002-2875-9567]{Martin}{Perešíni}

\address{
    Security@FIT, Brno University of Technology, Czech Republic
}

\email{\{ifirc, ilicka, istanek, malinka, iperesini\}@fit.vut.cz}


\usepackage[table]{xcolor}
\usepackage{comment}
\usepackage{amsmath}
\usepackage{amssymb}    
\usepackage{braket}     
\usepackage{multirow}   
\usepackage{multicol}   
\usepackage{svg}        
\usepackage{tabularx}   
\usepackage{tikz}
\usepackage{subcaption}
\usepackage[skip=9pt]{caption}
\usepackage{amsmath}
\usetikzlibrary{positioning,fit,arrows.meta}

\renewcommand{\paragraph}[1]{\vspace{0.5ex}\noindent\textbf{#1}\hspace{0.75em}}

\begin{document}

\maketitle

\begin{abstract}
We present \textbf{\texttt{SpAArSIST}}, a deployment-oriented refinement of the widely used \texttt{AASIST} graph pooling backend for self-supervised learning (SSL) based anti-spoofing. Motivated by redundant operations in public implementations, we replace learned pooling and stack-node attention with explicit, lightweight choices: separate train and inference graph pooling ratios $(k_{\mathrm{tr}},k_{\mathrm{inf}})$, magnitude-based node scoring, and mean aggregation of graph nodes. The best overall configuration (rank~1) cuts backend compute by 20.7\% (195.045M $\rightarrow$ 154.706M MACs) and model size by 4.1\% (611.8k $\rightarrow$ 586.4k params), while improving out-of-domain robustness on In-the-Wild to 2.82\% EER and 0.078 minDCF (from 4.64\% and 0.133) and remaining competitive on ASVspoof~5. We further provide a composite selection score that summarizes accuracy, calibration, and compute to support balanced deployment-oriented model choice.
\end{abstract}

\section{Introduction}
Speech spoofing and deepfake generation methods continue to improve in quality and diversity, increasing the demand for robust, deployable spoofing countermeasures~\cite{speech-biometric-practicalattack_firc22,voiceAssitants,humanBiosig}.  
Modern detection pipelines typically combine a pretrained self-supervised learning front-end (e.g., \texttt{XLS-R}~\cite{wav2vec2-xls-r}, \texttt{Wav2Vec2.0}~\cite{wav2vec2} or \texttt{WavLM}~\cite{wavlm}), with a learnable pooling backend (e.g., \texttt{AASIST}~\cite{aasist,tak22_odyssey}, \texttt{MHFA}~\cite{mhfa-pooling-architecture,ca-mhfa}, \texttt{SLS}~\cite{sls-pooling-architecture_zhang24}, or a simple mean pooling~\cite{rohdin24_asvspoof}) that aggregates frame-level representations into an utterance-level embedding for binary classification~\cite{evalFramework,asvspoof5-challenge}.

Within pooling backends, \texttt{AASIST}~\cite{aasist, tak22_odyssey}
models spectro-temporal relationships through graph-based operations and promotes strong class separation in the embedding space.  
Proposed early in the deepfake detection literature and demonstrating consistently strong results, \texttt{AASIST} has become one of the most widely adopted pooling modules for spoofing detection and related tasks~\cite{asvspoof5-challenge}.

While the concept remains very effective in anti-spoofing, we observed that the commonly used public implementations of \texttt{AASIST}\footnote{\scriptsize\url{https://github.com/TakHemlata/SSL_Anti-spoofing},\\\url{https://github.com/clovaai/aasist}}
contain redundant or weakly conditioned operations.  
These design choices increase computational cost without providing proportional representational benefit.  
Thus, we introduce \textbf{\texttt{SpAArSIST}}\footnote{\scriptsize\url{https://github.com/Security-FIT/SpAArSIST}}:
a systematic, efficiency-oriented refinement of \texttt{AASIST}. Based on systematic variation, we strictly isolate each optimization's contribution.

Unlike recent \texttt{AASIST} extensions that improve accuracy by adding architectural components~\cite{aasist3-pooling-architecture_borodin24,scalable-aasist-pooling-architecture_viakhirev25}, we revisit the baseline implementation and ask how far simplification can go without sacrificing performance. We find that several backend operations in widely used codebases contribute limited benefit relative to their cost, and we replace them with explicit, lightweight alternatives that make the pruning and aggregation behavior easier to control. In the \texttt{XLS-R}+backend pipeline, \texttt{SpAArSIST} reduces model size by 4.1\% (611.8k $\rightarrow$ 586.4k params) and backend compute by 20.7\% (195.045M $\rightarrow$ 154.706M MACs). 
At the same time, it improves robustness on In-the-Wild, reducing EER from 4.64\% to 2.82\% and minDCF from 0.133 to 0.078, while remaining competitive in-domain on ASVspoof~5. To support deployment-oriented comparison across variants, we also report a two-track composite score that combines accuracy, calibration, and compute into a single selection criterion.

\noindent The \textbf{novel contributions} include:
\begin{itemize}
\item We propose \textbf{SpAArSIST}, a deployment-oriented redesign of the \texttt{AASIST} backend that streamlines graph pooling and pruning to reduce compute while improving generalization and calibration.
\item We introduce separate \textbf{train-time and inference-time pooling ratios} $(k_{\text{tr}}, k_{\text{inf}})$ and quantify the trade-off between compute and accuracy under domain shift. 
\item We replace attention-heavy components with \textbf{lightweight node scoring and aggregation} proxies and validate the substitutions via targeted ablations using discrimination and calibration metrics. 
\end{itemize}

\section{Related Work}
\textbf{Anti-spoofing countermeasures:} 
Early countermeasures relied on handcrafted acoustic features and shallow classifiers~\cite{BUT185123}.
Deep learning approaches later improved separation by learning robust representations directly from waveforms or spectrograms, often aided by data augmentation~\cite{rawboost} and challenge protocols~\cite{asvspoof5-challenge-dataset-design, asvspoof21-challenge, add2023-challenge}.

Deep learning approaches have been exclusively adapted as self-supervised learning (SSL) encoders (e.g., \texttt{Wav2Vec2.0}~\cite{wav2vec2}, \texttt{XLS-R}~\cite{wav2vec2-xls-r}, \texttt{WavLM}~\cite{wavlm}), which improve representation quality and cross-condition generalization~\cite{tak22_odyssey}.
They have been shown to outperform handcrafted features using large amounts of training data~\cite{asvspoof5-challenge}.

These encoders are paired with aggregation techniques (e.g., \texttt{MHFA}~\cite{mhfa-pooling-architecture,ca-mhfa}, \texttt{SLS}~\cite{sls-pooling-architecture_zhang24}), including graph-based pooling (e.g., \texttt{AASIST}~\cite{aasist}).
Graph-based backends model relationships between temporal and spectral structures, thereby exposing spoofing artifacts that manifest as relational inconsistencies rather than localized distortions.
AASIST-style models have become a widely adopted baseline~\cite{asvspoof5-challenge}, which makes them a strong contender for controlled optimization.

\textbf{AASIST Optimization:}
Motivated by its adoption, a few studies have emerged to improve the \texttt{AASIST} architecture~\cite{aasist3-pooling-architecture_borodin24,scalable-aasist-pooling-architecture_viakhirev25}.
Even the original \texttt{AASIST}~\cite{aasist} work reduces the larger \texttt{AASIST} to a lightweight version, \texttt{AASIST-L}, using a population-based training algorithm, trading size 
for performance. 
Borodin et al.~\cite{aasist3-pooling-architecture_borodin24} improve \texttt{AASIST} by introducing Kolmogorov-Arnold networks, additional layers, encoders, and pre-emphasis techniques.
The authors improve the t-DCF by more than 50\% relative to the baseline \texttt{AASIST} on the ASVspoof5 dataset~\cite{asvspoof5-challenge}.
Viakhirev et al.~\cite{scalable-aasist-pooling-architecture_viakhirev25} refined the \texttt{AASIST} architecture by replacing its bespoke graph-attention and heuristic max-fusion layers with standardized multi-head attention and a trainable, soft-fusion module; achieving a 7.66\% EER on ASVspoof 5.


Unlike works that add new front ends or heavier backbones, we optimize the \texttt{AASIST} implementation itself by replacing redundant graph operations and attention-heavy blocks with lightweight alternatives. This keeps the front end fixed while \textbf{cutting backend compute by 20.7\%} and improving out-of-domain robustness on In-the-Wild (EER 4.64\% $\rightarrow$ 2.82\%), as detailed in \autoref{sec:results}.


\section{Methodology}

\texttt{AASIST}~\cite{tak22_odyssey} follows a graph-based pooling backend on top of an SSL front-end. For an input utterance, the processing can be summarized as:
\begin{enumerate}
  \item \textbf{Front-end feature extraction:} a pretrained SSL encoder produces frame-level representations.
  \item \textbf{Graph construction:} the utterance is mapped to graph nodes (frames or regions), typically forming parallel spectral and temporal views.
  \item \textbf{Graph interaction:} graph-attention blocks propagate information between nodes to model longer-range spectro-temporal relations that can expose spoofing artifacts.
  \item \textbf{Node scoring and pooling:} nodes are assigned importance scores and a top-$k$ subset is retained to reduce later computation and focus on salient regions.
  \item \textbf{Graph readout (stack node):} A dedicated stack node aggregates information from the retained nodes to form an utterance-level embedding.
  \item \textbf{Classification:} a lightweight prediction head maps the utterance embedding to bona fide vs.\ spoof.
\end{enumerate}

Conceptually, \texttt{AASIST} builds two complementary graph views of the same utterance, one emphasizing temporal structure and one emphasizing spectral structure, then couples them through heterogeneous graph attention so cues from either view can influence the final decision. The stack node serves as a global summary token: it is updated by attending to the retained nodes, turning a variable-size set of node embeddings into a fixed-dimensional utterance representation.

Finally, the node-scoring and pooling stage serves as a competitive selection mechanism. By keeping only the highest-scoring nodes (top-$k$), \texttt{AASIST} suppresses less informative regions and concentrates subsequent graph interaction and readout on the most suspicious parts of the signal.


\subsection{SpAArSIST changes and motivation}
SpAArSIST targets two coupled goals: reducing backend compute and improving operational performance under domain shift.
The main modifications are summarized below.

\paragraph{(1) Train-time vs inference-time node pooling $(k_\text{tr}, k_\text{inf})$.}
Graph operations scale with the number of retained nodes, so reducing the node pooling ratio directly reduces the number of backend MACs.
We study a baseline-compatible regime ($0.5$ retention) and a reduced regime ($0.3$, $0.1$ retention) that aggressively shrinks later graph operations.

We expose a top-$k$ control at both training and inference.
This enables two distinct behaviors:
\begin{itemize}
  \item \textbf{Matched sparsity}: $k_{\mathrm{tr}} = k_{\mathrm{inf}}$, i.e., inference uses the same retention ratio as training.
  \item \textbf{Inference-only sparsity}: $k_\text{inf} < k_\text{tr}$, which can further reduce runtime.
\end{itemize}
Throughout, \emph{sparsity} denotes the retained-node fraction set by top-$k$ pooling (lower retention means a sparser graph), not weight or activation sparsity.



\paragraph{(2) GraphPool scoring simplification via magnitude proxy.}
In \texttt{AASIST}, GraphPool first assigns each node an importance score and then retains the top-$k$ nodes for subsequent graph interaction and readout. The standard implementation uses a learned scorer, typically a linear projection followed by a sigmoid,
\begin{equation}
s_i = \sigma(\mathbf{w}^\top \mathbf{n}_i + b),
\end{equation}
which introduces additional parameters and a saturating nonlinearity whose scale can implicitly change the pooling behavior.

In \texttt{SpAArSIST}, we replace this learned path with a parameter-free, compute-light proxy that scores each node by its feature magnitude (denoted as \textbf{Mag} in \autoref{tab:main}),
\begin{equation}
s_i = \lVert \mathbf{n}_i \rVert_2,
\end{equation}
So, pooling reduces to ranking nodes by energy in the learned representation and selecting the top-$k$ indices. Since GraphPool only depends on the ordering of $\{s_i\}$, any strictly monotonic transform yields the same selection; therefore, we can equivalently use the squared norm to avoid the square root,
\begin{equation}
s_i \propto \lVert \mathbf{n}_i \rVert_2^2 = \sum_{d} n_{i,d}^2.
\end{equation}

This substitution removes the scorer parameters $(\mathbf{w}, b)$ and eliminates the sigmoid, while preserving the key role of the scoring stage, which is to focus computation on a small subset of salient nodes. We do not claim magnitude is universally optimal. Instead, we treat it as an explicit, interpretable proxy whose effect can be cleanly isolated in ablations, especially when varying the pooling ratio $k$ and evaluating generalization under domain shift.

\paragraph{(3) Stack-node aggregation simplification motivated by temperature.}
Let $\{\mathbf{n}_i\}_{i=1}^{M}$ denote the $M$ retained nodes after GraphPool. In \texttt{AASIST}, the stack node forms an utterance embedding by attention-weighted aggregation of the retained nodes. Abstracting away implementation details, this can be written as
\begin{align}
\boldsymbol{\alpha}(\tau) &= \operatorname{softmax}(\mathbf{g}/\tau) \\
\mathbf{g} &= [g(\mathbf{n}_1),\ldots,g(\mathbf{n}_M)]^{\mathsf{T}} \\
\mathbf{z}(\tau) &= \sum_{i=1}^{M}\alpha_i(\tau)\,\mathbf{n}_i
\end{align}
where $g(\cdot)$ is a learned compatibility score (e.g., a dot product with the current stack state) and $\tau>0$ is the softmax temperature. The role of $\tau$ is to control selectivity: small $\tau$ yields peaky weights, large $\tau$ yields flat weights.

\textbf{Mean-based stack update.}
We observed that common public implementations use very large temperatures (e.g., $\tau=100$), which makes $\alpha_i(\tau)$ close to uniform and causes the stack update to behave like an (almost) unweighted average. This follows directly from
\begin{equation}
\lim_{\tau\rightarrow\infty}\alpha_i(\tau)=\frac{1}{M}
\qquad \Rightarrow \qquad
\mathbf{z}_{\text{mean}}=\frac{1}{M}\sum_{i=1}^{M}\mathbf{n}_i.
\end{equation}
Motivated by this effective behavior, we replace the attention aggregation with an explicit mean (\textbf{Mean} in \autoref{tab:main}):
\begin{equation}
\mathbf{z}_{\text{mean}}=\frac{1}{M}\sum_{i=1}^{M}\mathbf{n}_i,
\end{equation}
which removes the score computation $g(\cdot)$ and the softmax normalization in the stack update while preserving the averaging behavior induced by high $\tau$.

\textbf{Temperature reduction.}
As an alternative to removing attention, we keep the same aggregation form but lower the temperature to restore non-uniform weighting:
\begin{equation}
\mathbf{z}(\tau_{\text{low}})=\sum_{i=1}^{M}\alpha_i(\tau_{\text{low}})\,\mathbf{n}_i,
\qquad \tau_{\text{low}}<\tau_{\text{base}}, (\tau_{\text{base}} = 100).
\end{equation}
This treats $\tau$ as an explicit control knob, allowing us to test whether more selective stack aggregation improves operating-point robustness and calibration.

\section{Experiment Setup}
\label{sec:experimental_framework}

\subsection{Architectural Backbone and Feature Pipeline}
Our experiments utilize a unified Self-Supervised Learning (SSL) frontend. We employ the \texttt{Wav2Vec2.0 XLS-R (300M)} front-end to generate frame-level representations. These features are subsequently integrated by an \texttt{AASIST}/\texttt{SpAArSIST} pooling layer, resulting in a fixed-size utterance embedding $h \in \mathbb{R}^{1024}$. 

Models are trained with a fixed pooling ratio $k_{tr}$ using the baseline backend. At evaluation, we optionally replace the pooling scorer and stack readout with magnitude scoring and mean aggregation, and we vary $k_{inf}$ to test inference-time sparsity.


\subsection{Data Resources and Training Protocol}
The primary training and development are based on the ASVspoof~5 Track 1 (ASV5) corpus~\cite{asvspoof5-challenge-dataset-design}. To test the cross-domain robustness of our models, we evaluate performance on both the ASV5 evaluation set (in-domain) and the In-the-Wild (ITW) dataset (out-of-domain)~\cite{muller2022does}.

We minimize softmax cross-entropy using the Adam optimizer with a learning rate of $1 \times 10^{-4}$. The training process follows a dual-stage schedule: \textit{1) Initial Phase:} 10 epochs with frozen extractor, and a batch size of 64, selecting the optimal checkpoint based on development-set (ASV5) EER. \textit{2) Refinement Phase:} 5 epochs of end-to-end fine-tuning with a reduced batch size of 32.

\paragraph{Data Augmentations:} To ensure generalizability, we apply a robust augmentation suite including Trim Starting Silence ($p=0.5$), Time Masking ($p=0.3$), Mu-law Companding ($p=0.3$), RawBoost (LnL-ISD) ($p=0.3$), and Noise Filtering ($p=0.3$).

\subsection{Evaluation Benchmarks}
Performance is quantified using standard ASVspoof metrics derived from the spoof logit~\cite{asvspoof5-challenge}:
\begin{itemize}
    \item \textbf{Discrimination:} Evaluated via Equal Error Rate (EER) and minimum Detection Cost Function (minDCF).
    \item \textbf{Calibration:} Assessed using actual DCF (actDCF), the log-likelihood ratio cost ($C_{llr}$), and Expected Calibration Error (ECE).
    \item \textbf{Computational Efficiency:} Measured by backend multiply-accumulate operations in millions (BE M-MACs) and backend-only forward latency (Proc Lat, ms, batch{=}1, \texttt{XLS-R} excluded, NVIDIA RTX A5000 24GB).
\end{itemize}

\begin{table*}[t]
\centering
\caption{Representative \texttt{AASIST}/\texttt{SpAArSIST} configurations (\texttt{XLS-R}+backend) on in-domain ASVspoof5~\cite{asvspoof5-challenge} and out-of-domain In-the-Wild~\cite{muller2022does}. BE: backend MACs in millions; FE constant. Track scores follow Eq.~\eqref{eq:perf}--\eqref{eq:final} with min-max normalization (Eq.~\eqref{eq:norm}). \textit{Rank}: position by composite score $S$ (Eq.~\eqref{eq:final}), lower is better. Baseline marked by separators.}
\label{tab:main}
\resizebox{\textwidth}{!}{%
\begin{tabular}{l c c c c c c c c c c c c c c c}
\toprule
\multirow{2}{*}{\textbf{ID}} &
\multicolumn{4}{c}{\textbf{Architecture}} &
\multicolumn{2}{c}{\textbf{System}} &
\multicolumn{4}{c}{\textbf{ASVspoof 5 (ASV5)}} &
\multicolumn{4}{c}{\textbf{In-the-Wild (ITW)}} &
\multicolumn{1}{c}{\textbf{Ranking}} \\
\cmidrule(lr){2-5} \cmidrule(lr){6-7} \cmidrule(lr){8-11} \cmidrule(lr){12-15} \cmidrule(lr){16-16}
& \textbf{$k_{\text{tr}}$} & \textbf{$k_{\text{inf}}$} & \textbf{Mag} & \textbf{Mean}
& \textbf{BE M-MACs} & \textbf{Proc Lat}
& \textbf{EER (\%)} & \textbf{Cllr} & \textbf{actDCF} & \textbf{minDCF}
& \textbf{EER (\%)} & \textbf{Cllr} & \textbf{actDCF} & \textbf{minDCF}
& \textbf{Rank} \\
\midrule
AST-03-01-Mag     & 0.3 & 0.1 & \checkmark &            & 154.706 & 7.468 & 5.05 & 0.481 & 0.156 & 0.146 & 2.82 & 0.374 & 0.081 & 0.078 & \textbf{1} \\
AST-03-01-MagMean & 0.3 & 0.1 & \checkmark & \checkmark & 154.706 & 7.448 & 5.05 & 0.481 & 0.156 & 0.146 & 2.82 & 0.374 & 0.081 & 0.078 & \textbf{1} \\
AST-03-03-Mag     & 0.3 & 0.3 & \checkmark &            & 167.087 & 7.830 & 5.08 & 0.442 & \textbf{0.151} & 0.147 & 2.82 & \textbf{0.349} & \textbf{0.079} & 0.078 & 3 \\
AST-03-03-MagMean & 0.3 & 0.3 & \checkmark & \checkmark & 167.087 & 7.532 & 5.08 & 0.442 & \textbf{0.151} & 0.147 & 2.82 & \textbf{0.349} & \textbf{0.079} & 0.078 & 3 \\
AST-03-01-Base    & 0.3 & 0.1 &            &            & 154.720 & 7.139 & 5.22 & 0.633 & 0.198 & 0.151 & 3.15 & 0.462 & 0.109 & 0.088 & 5 \\
AST-03-01-Mean    & 0.3 & 0.1 &            & \checkmark & 154.720 & 7.378 & 5.22 & 0.633 & 0.198 & 0.151 & 3.15 & 0.462 & 0.109 & 0.088 & 5 \\
AST-03-03-Base    & 0.3 & 0.3 &            &            & 167.102 & 7.414 & 5.25 & 0.671 & 0.194 & 0.152 & 3.10 & 0.462 & 0.102 & 0.086 & 7 \\
AST-03-03-Mean    & 0.3 & 0.3 &            & \checkmark & 167.102 & 7.646 & 5.25 & 0.671 & 0.194 & 0.152 & 3.10 & 0.462 & 0.102 & 0.086 & 7 \\
AST-01-01-Base    & 0.1 & 0.1 &            &            & 154.720 & \underline{8.870} & 4.66 & 0.862 & 0.218 & 0.134 & 4.21 & 0.951 & 0.201 & 0.122 & 9 \\
AST-05-03-Mag     & 0.5 & 0.3 & \checkmark &            & 167.087 & 7.748 & 4.54 & 1.209 & 0.253 & 0.131 & 12.49 & 0.936 & 0.375 & 0.206 & 10 \\
AST-01-01-Mag     & 0.1 & 0.1 & \checkmark &            & 154.706 & 7.900 & 4.98 & 1.113 & 0.232 & 0.142 & \underline{25.40} & 0.976 & 0.864 & 0.389 & 11 \\
AST-05-03-Base    & 0.5 & 0.3 &            &            & 167.102 & 7.333 & 4.55 & 1.703 & 0.332 & 0.131 & 3.28 & 2.091 & 0.386 & 0.095 & 12 \\
AST-05-01-Mag     & 0.5 & 0.1 & \checkmark &            & 154.706 & 7.219 & 4.61 & 1.879 & 0.312 & 0.133 & 14.92 & 1.405 & 0.393 & 0.253 & 13 \\
AST-05-05-Mean    & 0.5 & 0.5 &            & \checkmark & 154.329 & 4.396 & 5.31 & 0.413 & 1.617 & 0.153 & 2.38 & 0.447 & 0.749 & \textbf{0.063} & 14 \\
AST-05-01-MagMean & 0.5 & 0.1 & \checkmark & \checkmark & \textbf{153.209} & 4.425 & 5.64 & 0.561 & 1.686 & 0.163 & 6.72 & 0.568 & 1.029 & 0.149 & 15 \\
AST-05-05-Mag     & 0.5 & 0.5 & \checkmark &            & 195.027 & 7.589 & 4.54 & 0.838 & 0.203 & 0.131 & 9.44 & 0.712 & 0.316 & 0.168 & 16 \\
AST-05-03-MagMean & 0.5 & 0.3 & \checkmark & \checkmark & 153.717 & 4.488 & 5.39 & 0.503 & 1.894 & 0.156 & 2.96 & 0.511 & 1.551 & 0.084 & 17 \\
\cmidrule(lr){1-16}
AST-05-05-Base    & 0.5 & 0.5 &            &            & \underline{195.045} & 7.168 & \textbf{4.49} & 1.261 & 0.272 & \textbf{0.129} & 4.64 & 1.407 & 0.291 & 0.133 & 18 \\
\cmidrule(lr){1-16}
AST-01-01-MagMean & 0.1 & 0.1 & \checkmark & \checkmark & \textbf{153.209} & 5.389 & 6.35 & \textbf{0.389} & \underline{1.900} & 0.184 & 19.42 & 0.553 & 1.899 & \underline{0.430} & 19 \\
AST-05-01-Base    & 0.5 & 0.1 &            &            & 154.720 & 6.971 & 4.71 & \underline{2.287} & 0.404 & 0.136 & 2.88 & \underline{2.957} & 0.506 & 0.083 & 20 \\
AST-05-05-MagMean & 0.5 & 0.5 & \checkmark & \checkmark & 154.311 & \textbf{4.324} & 5.27 & 0.472 & \underline{1.900} & 0.153 & \textbf{2.26} & 0.485 & 1.876 & 0.064 & 21 \\
AST-01-01-Mean    & 0.1 & 0.1 &            & \checkmark & 153.222 & 5.323 & \underline{7.76} & 0.858 & \underline{1.900} & \underline{0.204} & 3.67 & 0.747 & \underline{1.900} & 0.069 & \underline{22} \\
\bottomrule
\end{tabular}%
}

\end{table*}

\subsection{Two-track composite score for ranking}
To compare systems across compute, discrimination, and reliability, we define a two-track composite score.

For each ``lower is better'' metric $x$, we apply inverted min-max normalization over the set of compared systems:
\begin{equation}
\mathcal{N}(x) = \frac{x_{\max} - x}{x_{\max} - x_{\min}} \in [0,1].
\label{eq:norm}
\end{equation}

\noindent \textbf{Performance track.} $\mathcal{M}_{\mathrm{perf}}$ includes the $\mathrm{minDCF}_{\mathrm{ASV5}}$, $\mathrm{Cllr}_{\mathrm{ASV5}}$,
$\mathrm{minDCF}_{\mathrm{ITW}}$, $\mathrm{Cllr}_{\mathrm{ITW}}$, and  BE MACs.
We compute a geometric mean to penalize weak components:
\begin{equation}
S_{\mathrm{perf}} =
\left(
\prod_{m \in \mathcal{M}_{\mathrm{perf}}} \mathcal{N}(m)
\right)^{\frac{1}{|\mathcal{M}_{\mathrm{perf}}|}}.
\label{eq:perf}
\end{equation}

\noindent \textbf{Reliability track.} $\mathcal{M}_{\mathrm{rel}}$ includes the  $\mathrm{Gap}_{\mathrm{ASV5}}$, $\mathrm{Gap}_{\mathrm{ITW}}$,$\mathrm{ECE}_{\mathrm{ASV5}}$, $\mathrm{ECE}_{\mathrm{ITW}}$, $\mathrm{actDCF}_{\mathrm{ASV5}}$, $\mathrm{actDCF}_{\mathrm{ITW}}$ metrics, where gap:
\begin{equation}
\mathrm{Gap} = \mathrm{actDCF} - \mathrm{minDCF}.
\label{eq:gap}
\end{equation}
We compute the geometric mean over $\mathcal{M}_{\mathrm{rel}}$ in similar manner (\autoref{eq:perf}), resulting in $S_{\mathrm{rel}}$.

\noindent \textbf{Final score (harmonic mean).}
Finally, we combine both tracks using a harmonic mean to punish systems that are strong in only one track:
\begin{equation}
S =
\frac{2 S_{\mathrm{perf}} S_{\mathrm{rel}}}{S_{\mathrm{perf}} + S_{\mathrm{rel}}}.
\label{eq:final}
\end{equation}

\noindent
\autoref{tab:main} reports only the final rank derived from the underlying $S$ metrics.

\section{Results}
\label{sec:results}

\autoref{tab:main} reports representative \texttt{AASIST}/\texttt{SpAArSIST} configurations, metrics, and composite scores. The top-ranked systems pair magnitude-based scoring with more aggressive pruning ($k_{\mathrm{tr}}=0.3$, $k_{\mathrm{inf}}=0.1$), remaining competitive on ASVspoof~5 while improving robustness on ITW.

\autoref{tab:pooling_baselines_compact} compares pooling backends under the same \texttt{XLS-R} front end and training pipeline. \texttt{SpAArSIST} reduces backend MACs and improves ITW performance relative to \texttt{AASIST}, while mean and MHFA pooling are less competitive under at least one evaluation condition.

\paragraph{Overall Performance.}
In-domain on ASVspoof~5, our best overall system \texttt{AST-03-01-Mag} (rank~1) remains close to the \texttt{AASIST} baseline in discrimination, with EER 5.05\% vs.\ 4.49\% and minDCF 0.146 vs.\ 0.129, while substantially improving calibration-oriented measures, with actDCF 0.156 vs.\ 0.272 and $C_{llr}$ 0.481 vs.\ 1.261. 

On the out-of-domain ITW evaluation set, the same configuration provides clear gains across both discrimination and calibration, reducing EER from 4.64\% to 2.82\%, minDCF from 0.133 to 0.078, actDCF from 0.291 to 0.081, and $C_{llr}$ from 1.407 to 0.374, while also reducing backend compute from 195.045M to 154.706M MACs. Across the ablations, lower train-time node retention ($k_{\mathrm{tr}}=0.3$) is repeatedly associated with stronger ITW results, and magnitude-based node scoring is most beneficial in this low-retention regime. Mean-only stack aggregation can improve efficiency and sometimes improve discrimination, including cases with very low ITW EER and minDCF, but it more often degrades actDCF and threshold-transfer behavior under shift, which is reflected in lower reliability scores and therefore weaker overall ranking under the harmonic mean criterion (\autoref{eq:final}).

\paragraph{Computational Efficiency.}
\texttt{SpAArSIST} yields consistent backend efficiency gains. For the $k_{\mathrm{tr}}=0.3$ configurations, backend compute drops by 20.7\% from 195.045M to 154.706M MACs. While the front-end (\texttt{XLS-R (300M)}) still dominates total system MACs, reducing backend graph operations lowers the portion of computation that is most sensitive to memory traffic and kernel overhead, which is beneficial for deployment.

Across ablations, additional inference-time sparsification ($k_{\mathrm{inf}}<k_{\mathrm{tr}}$) further reduces backend compute with limited impact on accuracy in the high-performing regimes. In our results, the largest MAC reductions are driven by pruning controls and lightweight node scoring, while mean-based aggregation provides a smaller latency benefit and primarily affects efficiency.

\paragraph{Softmax Temperature.}
Lowering the stack-node softmax temperature does not improve over the default high-temperature setting ($\tau=100$) in our main results. While reduced temperatures ($\tau\in\{1,5,20\}$) can yield competitive out-of-domain performance for mean-based aggregation, the gains are not as consistent as those of the best default-temperature configurations. For example, \texttt{AST-03-01-Mean} at $\tau=5$ reaches ITW EER 2.98\% and minDCF 0.082, but remains behind the strongest default-$\tau$ systems in overall stability. In contrast, combining low temperatures with graph pruning based on the magnitude proxy is unstable and can severely degrade both in-domain and out-of-domain performance (ASV5 EER 16.66\%; ITW EER 35.99\%). Overall, the temperature sweep supports the interpretation that the original stack update already operates close to a mean-like regime, and that explicit mean aggregation is a safer way to capture this behavior than tuning $\tau$.

\begin{table}[t]
\centering
\caption{Comparison of pooling backends under the same XLS-R (300M) frontend and training recipe.
Backend MACs and Params exclude frontend compute and parameters (constant).} 
\label{tab:pooling_baselines_compact}
\scriptsize
\setlength{\tabcolsep}{3.0pt}
\renewcommand{\arraystretch}{1.05}
\resizebox{\linewidth}{!}{%
\begin{tabular}{lcc|c|cc}
\toprule
\multirow{2}{*}{System} & \multirow{2}{*}{\shortstack{BE Params\\(k) $\downarrow$}} & \multirow{2}{*}{\shortstack{BE M-MACs\\ $\downarrow$}} & \multirow{2}{*}{\shortstack{ASVspoof 5\\EER (\%) $\downarrow$}} & \multicolumn{2}{c}{ITW} \\
\cmidrule(lr){5-6}
& & & & EER (\%) $\downarrow$ & actDCF $\downarrow$ \\ 
\midrule
AASIST (base) &
611.8 &
195.05 &
4.49 &
4.64 &
0.291 \\

Mean pooling &
0.0 &
0.31 &
4.72 &
4.29 &
0.496 \\

MHFA~\cite{ca-mhfa} &
4461.9 &
99.49 &
6.05 &
4.98 &
0.350 \\

\rowcolor{black!8}
\textbf{AST-03-01-Mag} &
586.4 &
154.71 &
5.05 &
\textbf{2.82} &
0.081 \\
\bottomrule
\end{tabular}%
}
\vspace{-2em}
\end{table}

\section{Conclusion}
We proposed \texttt{SpAArSIST}, a deployment-oriented simplification of the \texttt{AASIST} graph backend in an \texttt{XLS-R} pipeline. Our results indicate that several commonly used graph components are not essential for robust spoofing detection: the stack-node attention behaves close to mean aggregation, and temperature tuning does not yield consistent gains, so the attention update can be replaced by a simple mean during inference. Together with explicit train-time and inference-time pooling control and compute-light node scoring, these simplifications reduce the number of backend graph operations and the amount of compute, as evidenced by the decrease in backend multiply-accumulate (BE M-MAC) operations. For our best overall system (rank~1, \texttt{AST-03-01-Mag}), out-of-domain performance on In-the-Wild improves to 2.82\% EER and 0.078 minDCF (from 4.64\% and 0.133 for the baseline), while remaining competitive in-domain on ASVspoof~5.

\ifcameraready
\section{Acknowledgments}
     This work was partially supported by the Brno University of Technology (internal project FIT-S-23-8151) and the Ministry of Education, Youth and Sports of the Czech Republic through the e-INFRA CZ (ID:90254).
\fi

\section{Generative AI Use Disclosure}
During the preparation of this work, the authors used Generative AI Models (specifically Google Gemini, ChatGPT, and Grammarly) for language editing and text refinement. The authors reviewed and edited the output as needed and take full responsibility for the publication's content.

\bibliographystyle{IEEEtran}
\bibliography{mybib}

@article{wavlm,
    title=          {WavLM: Large-Scale Self-Supervised Pre-Training for Full Stack Speech Processing},
    volume=         {16},
    issn=           {1941-0484},
    doi=            {10.1109/jstsp.2022.3188113},
    number=         {6},
    journal=        {IEEE Journal of Selected Topics in Signal Processing},
    publisher=      {Institute of Electrical and Electronics Engineers (IEEE)},
    author=         {Chen, Sanyuan and Wang, Chengyi and Chen, Zhengyang and Wu, Yu and Liu, Shujie and Chen, Zhuo and Li, Jinyu and Kanda, Naoyuki and Yoshioka, Takuya and Xiao, Xiong and Wu, Jian and Zhou, Long and Ren, Shuo and Qian, Yanmin and Qian, Yao and Wu, Jian and Zeng, Michael and Yu, Xiangzhan and Wei, Furu},
    year=           {2022},
    month=          {July},
    day=            {04},
    pages=          {1505–1518},
    howpublished=   {online},
    url=            {http://dx.doi.org/10.1109/JSTSP.2022.3188113},
    cited=          {01-05-2024}
}

@inproceedings{speech-biometric-practicalattack_firc22,
author = {Firc, Anton and Malinka, Kamil},
title = {The dawn of a text-dependent society: deepfakes as a threat to speech verification systems},
year = {2022},
isbn = {9781450387132},
publisher = {Association for Computing Machinery},
address = {New York, NY, USA},
url = {https://doi.org/10.1145/3477314.3507013},
doi = {10.1145/3477314.3507013},
pages = {1646–1655},
numpages = {10},
location = {Virtual Event},
series = {SAC '22}
}

@misc{wav2vec2,
      title={wav2vec 2.0: A Framework for Self-Supervised Learning of Speech Representations}, 
      author={Alexei Baevski and Henry Zhou and Abdelrahman Mohamed and Michael Auli},
      year={2020},
      eprint={2006.11477},
      archivePrefix={arXiv},
      primaryClass={cs.CL},
      url={https://arxiv.org/abs/2006.11477}, 
}

@misc{wav2vec2-xls-r,
      title={XLS-R: Self-supervised Cross-lingual Speech Representation Learning at Scale}, 
      author={Arun Babu and Changhan Wang and Andros Tjandra and Kushal Lakhotia and Qiantong Xu and Naman Goyal and Kritika Singh and Patrick von Platen and Yatharth Saraf and Juan Pino and Alexei Baevski and Alexis Conneau and Michael Auli},
      year={2021},
      eprint={2111.09296},
      archivePrefix={arXiv},
      primaryClass={cs.CL},
      url={https://arxiv.org/abs/2111.09296}, 
}

@INPROCEEDINGS{rawboost,
  author={Tak, Hemlata and Kamble, Madhu and Patino, Jose and Todisco, Massimiliano and Evans, Nicholas},
  booktitle={ICASSP 2022 - 2022 IEEE International Conference on Acoustics, Speech and Signal Processing (ICASSP)}, 
  title={Rawboost: A Raw Data Boosting and Augmentation Method Applied to Automatic Speaker Verification Anti-Spoofing}, 
  year={2022},
  volume={},
  number={},
  pages={6382-6386},
  doi={10.1109/ICASSP43922.2022.9746213}
}

@INPROCEEDINGS{aasist,
  author={Jung, Jee-weon and Heo, Hee-Soo and Tak, Hemlata and Shim, Hye-jin and Chung, Joon Son and Lee, Bong-Jin and Yu, Ha-Jin and Evans, Nicholas},
  booktitle={ICASSP 2022 - 2022 IEEE International Conference on Acoustics, Speech and Signal Processing (ICASSP)}, 
  title={AASIST: Audio Anti-Spoofing Using Integrated Spectro-Temporal Graph Attention Networks}, 
  year={2022},
  volume={},
  number={},
  pages={6367-6371},
  doi={10.1109/ICASSP43922.2022.9747766}
}

@misc{asvspoof21-challenge,
      title={ASVspoof 2021: accelerating progress in spoofed and deepfake speech detection}, 
      author={Junichi Yamagishi and Xin Wang and Massimiliano Todisco and Md Sahidullah and Jose Patino and Andreas Nautsch and Xuechen Liu and Kong Aik Lee and Tomi Kinnunen and Nicholas Evans and Héctor Delgado},
      year={2021},
      eprint={2109.00537},
      archivePrefix={arXiv},
      primaryClass={eess.AS},
      url={https://arxiv.org/abs/2109.00537}, 
}

@misc{add2023-challenge,
      title={ADD 2023: the Second Audio Deepfake Detection Challenge}, 
      author={Jiangyan Yi and Jianhua Tao and Ruibo Fu and Xinrui Yan and Chenglong Wang and Tao Wang and Chu Yuan Zhang and Xiaohui Zhang and Yan Zhao and Yong Ren and Le Xu and Junzuo Zhou and Hao Gu and Zhengqi Wen and Shan Liang and Zheng Lian and Shuai Nie and Haizhou Li},
      year={2023},
      eprint={2305.13774},
      archivePrefix={arXiv},
      primaryClass={cs.SD},
      url={https://arxiv.org/abs/2305.13774}, 
}

@inproceedings{asvspoof5-challenge,
  title     = {ASVspoof 5: crowdsourced speech data, deepfakes, and adversarial attacks at scale},
  author    = {Xin Wang and Héctor Delgado and Hemlata Tak and Jee-weon Jung and Hye-jin Shim and Massimiliano Todisco and Ivan Kukanov and Xuechen Liu and Md Sahidullah and Tomi H. Kinnunen and Nicholas Evans and Kong Aik Lee and Junichi Yamagishi},
  year      = {2024},
  booktitle = {The Automatic Speaker Verification Spoofing Countermeasures Workshop (ASVspoof 2024)},
  pages     = {1--8},
  doi       = {10.21437/ASVspoof.2024-1},
}

@misc{asvspoof5-challenge-dataset-design,
      title={ASVspoof 5: Design, Collection and Validation of Resources for Spoofing, Deepfake, and Adversarial Attack Detection Using Crowdsourced Speech}, 
      author={Xin Wang and Héctor Delgado and Hemlata Tak and Jee-weon Jung and Hye-jin Shim and Massimiliano Todisco and Ivan Kukanov and Xuechen Liu and Md Sahidullah and Tomi Kinnunen and Nicholas Evans and Kong Aik Lee and Junichi Yamagishi and Myeonghun Jeong and Ge Zhu and Yongyi Zang and You Zhang and Soumi Maiti and Florian Lux and Nicolas Müller and Wangyou Zhang and Chengzhe Sun and Shuwei Hou and Siwei Lyu and Sébastien Le Maguer and Cheng Gong and Hanjie Guo and Liping Chen and Vishwanath Singh},
      year={2025},
      eprint={2502.08857},
      archivePrefix={arXiv},
      primaryClass={eess.AS},
      url={https://arxiv.org/abs/2502.08857}, 
}

@misc{mhfa-pooling-architecture,
      title={An attention-based backend allowing efficient fine-tuning of transformer models for speaker verification}, 
      author={Junyi Peng and Oldrich Plchot and Themos Stafylakis and Ladislav Mosner and Lukas Burget and Jan Cernocky},
      year={2022},
      eprint={2210.01273},
      archivePrefix={arXiv},
      primaryClass={eess.AS},
      url={https://arxiv.org/abs/2210.01273}, 
}

@inproceedings{sls-pooling-architecture_zhang24,
author = {Zhang, Qishan and Wen, Shuangbing and Hu, Tao},
title = {Audio Deepfake Detection with Self-Supervised XLS-R and SLS Classifier},
year = {2024},
isbn = {9798400706868},
publisher = {Association for Computing Machinery},
address = {New York, NY, USA},
url = {https://doi.org/10.1145/3664647.3681345},
doi = {10.1145/3664647.3681345},
pages = {6765–6773},
numpages = {9},
location = {Melbourne VIC, Australia},
series = {MM '24}
}

@inproceedings{aasist3-pooling-architecture_borodin24,
  title     = {{AASIST3: KAN-enhanced AASIST speech deepfake detection using SSL features and additional regularization for the ASVspoof 2024 Challenge}},
  author    = {Kirill Borodin and Vasiliy Kudryavtsev and Dmitrii Korzh and Alexey Efimenko and Grach Mkrtchian and Mikhail Gorodnichev and Oleg Y. Rogov},
  year      = {2024},
  booktitle = {{The Automatic Speaker Verification Spoofing Countermeasures Workshop (ASVspoof 2024)}},
  pages     = {48--55},
  doi       = {10.21437/ASVspoof.2024-8},
}

@misc{scalable-aasist-pooling-architecture_viakhirev25,
      title={Towards Scalable AASIST: Refining Graph Attention for Speech Deepfake Detection}, 
      author={Ivan Viakhirev and Daniil Sirota and Aleksandr Smirnov and Kirill Borodin},
      year={2025},
      eprint={2507.11777},
      archivePrefix={arXiv},
      primaryClass={cs.SD},
      url={https://arxiv.org/abs/2507.11777}, 
}

@article{muller2022does,
  title={Does audio deepfake detection generalize?},
  author={M{\"u}ller, Nicolas M and Czempin, Pavel and Dieckmann, Franziska and Froghyar, Adam and B{\"o}ttinger, Konstantin},
  journal={Interspeech},
  year={2022}
}

@inproceedings{tak22_odyssey,
  title     = {{Automatic Speaker Verification Spoofing and Deepfake Detection Using Wav2vec 2.0 and Data Augmentation}},
  author    = {Hemlata Tak and Massimiliano Todisco and Xin Wang and Jee-weon Jung and Junichi Yamagishi and Nicholas Evans},
  year      = {2022},
  booktitle = {{The Speaker and Language Recognition Workshop (Odyssey 2022)}},
  pages     = {112--119},
  doi       = {10.21437/Odyssey.2022-16},
}

@article{BUT185123,
  author="Anton {Firc} and Kamil {Malinka} and Petr {Hanáček}",
  title="Deepfakes as a threat to a speaker and facial recognition: an overview of tools and attack vectors",
  journal="Heliyon",
  year="2023",
  volume="9",
  number="4",
  pages="1--33",
  doi="10.1016/j.heliyon.2023.e15090",
  url="https://www.sciencedirect.com/science/article/pii/S2405844023022971"
}

@inproceedings{rohdin24_asvspoof,
  title     = {{BUT systems and analyses for the ASVspoof 5 Challenge}},
  author    = {Johan Rohdin and Lin Zhang and Plchot Oldřich and Vojtěch Staněk and David Mihola and Junyi Peng and Themos Stafylakis and Dmitriy Beveraki and Anna Silnova and Jan Brukner and Lukáš Burget},
  year      = {2024},
  booktitle = {{The Automatic Speaker Verification Spoofing Countermeasures Workshop (ASVspoof 2024)}},
  pages     = {24--31},
  doi       = {10.21437/ASVspoof.2024-4},
}

@INPROCEEDINGS{ca-mhfa,
  author={Peng, Junyi and Mošner, Ladislav and Zhang, Lin and Plchot, Oldřich and Stafylakis, Themos and Burget, Lukáš and Černocký, Jan},
  booktitle={ICASSP 2025 - 2025 IEEE International Conference on Acoustics, Speech and Signal Processing (ICASSP)}, 
  title={CA-MHFA: A Context-Aware Multi-Head Factorized Attentive Pooling for SSL-Based Speaker Verification}, 
  year={2025},
  volume={},
  number={},
  pages={1-5},
  doi={10.1109/ICASSP49660.2025.10889058}}

@InProceedings{voiceAssitants,
author="Malinka, Kamil
and Firc, Anton
and Ka{\v{s}}ka, Petr
and Lap{\v{s}}ansk{\'y}, Tom{\'a}{\v{s}}
and {\v{S}}andor, Oskar
and Homoliak, Ivan",
editor="Garcia-Alfaro, Joaquin
and Kozik, Rafa{\l}
and Chora{\'{s}}, Micha{\l}
and Katsikas, Sokratis",
title="Resilience of Voice Assistants to Synthetic Speech",
booktitle="Computer Security -- ESORICS 2024",
year="2024",
publisher="Springer Nature Switzerland",
address="Cham",
pages="66--84",
isbn="978-3-031-70879-4"
}

@Article{evalFramework,
author={Firc, Anton
and Malinka, Kamil
and Han{\'a}{\v{c}}ek, Petr},
title={Evaluation framework for deepfake speech detection: a comparative study of state-of-the-art deepfake speech detectors},
journal={Cybersecurity},
year={2025},
month={Aug},
day={01},
volume={8},
number={1},
pages={50},
issn={2523-3246},
doi={10.1186/s42400-024-00346-1},
url={https://doi.org/10.1186/s42400-024-00346-1}
}

@INPROCEEDINGS{humanBiosig,
  author={Prudký, Daniel and Firc, Anton and Malinka, Kamil},
  booktitle={2023 International Conference of the Biometrics Special Interest Group (BIOSIG)}, 
  title={Assessing the Human Ability to Recognize Synthetic Speech in Ordinary Conversation}, 
  year={2023},
  volume={},
  number={},
  pages={1-5},
  doi={10.1109/BIOSIG58226.2023.10346006}}

\end{document}